\documentstyle[aps]{revtex}
\begin{document}
\draft
\title {QED blue-sheet effects inside black holes}
\author{Lior M. Burko}
\address{
Department of Physics, Technion---Israel Institute of Technology,
32000 Haifa, Israel.}
\date{\today}
\maketitle

\begin{abstract}

The interaction of the 
unboundedly blue-shifted photons of the cosmic microwave background 
radiation with a physical object falling towards the inner horizon of 
a Reissner-Nordstr\"{o}m  
black hole is analyzed. To evaluate this interaction we consider the QED 
effects up to the second order in the perturbation expansion. We then 
extrapolate 
the QED effects up to a cutoff, which we introduce at the Planckian level. (Our 
results are not sensitive to the cutoff energy.) 
We find that the energy absorbed by an infalling observer is finite, and 
for typical parameters would not lead to a catastrophic heating. However, 
this interaction would almost certainly be fatal for a human being, or other 
living organism of similar size. On the other hand, we find 
that smaller objects may survive the interaction.   
Our results do not provide support to the idea that the Cauchy horizon is 
to be regarded as the boundary of spacetime.

\end{abstract}

\pacs{PACS numbers: 04.40.Nr, 04.70.Bw}

\nopagebreak

\narrowtext

\twocolumn 

\section{Introduction}
The question whether there is a boundary to spacetime  
is of fundamental interest for a more complete understanding of the 
structure of spacetime and spacetime singularities. 
It has been argued \cite{poissonandisrael} that the inner horizon of 
black holes (BHs) may be such a boundary. The inner horizon is a 3-surface of 
infinite 
blue shift. Namely, any ingoing radiation, even very mild and well-behaved 
in the {\em external} universe, is infinitely blue-shifted at the inner 
horizon \cite{penrose}, also known as the Cauchy horizon (CH). 

Even if spacetime continues beyond the CH, it is 
still not clear whether (extended) physical objects can traverse it without 
being destroyed by the infinite concentration of energy density and by the 
curvature singularity expected to lurk there. The main goal of this Paper is to 
study the following 
question within some (very) simplified model: Are extended objects 
completely destroyed by the infinite concentration of energy density 
expected to exist at the CH? To address this question we shall deal only 
with the {\it direct} electrodynamic effects of the blue sheet. 
Thus, we ignore the 
tidal effects on the infalling object. (These effects were 
found to be negligible at the early parts of the inner horizon singularity -- 
see Refs. \cite{ori1,ori2}.) Also, we ignore the 
influence of 
local curvature on the QED processes. Namely, we assume that the size 
of the object is smaller than the typical radius of spacetime 
curvature near the CH.   
However, there always exist gravitational perturbations which produce a
diverging curvature at the CH, which 
means that the radius of curvature vanishes there.
Our assumption is valid only if---for the sake of evaluating
the QED effects---we ignore the gravitational perturbations. 
The modification of the  interaction by the metric perturbations is 
obviously a non-linear
effect, as it is quadratic in the perturbation's amplitude. This non-linear
effect remains the subject of future research. We do not expect, however,
this non-linear effect to significantly alter the linear-order interaction.
The gravitational analogue of this problem---the object's interaction with
the divergent tidal forces---demonstrates this reasoning: As implied from the
analysis of Ref. \cite{ori2} on the strength of the CH singularity in
spinning BHs---where it has been demonstrated
that the non-linear gravitational interaction with an object
may be negligible---we do not expect
higher-order contributions to change the general picture significantly. 
Thus, in this work we restrict ourselves to linear
effects only, and take the background to be unperturbed.  

It is believed that 
when an astrophysical BH is formed (through a gravitational collapse 
process), after 
perturbations die off a Kerr BH is left. 
It turns out that mathematical analysis of perturbations in the Kerr 
BH is very complicated, due to the lack of spherical symmetry of the 
background \cite{ori3}. The Reissner-Nordstr\"{o}m (RN) BH is therefore 
often used as a toy model for the more physical Kerr BH. This model can be 
justified by the similarity between the inner causal structures of the 
two solutions and by similar blue-shift effects.  

In principle, any attempt to predict the fate of an object which tries 
to cross the CH 
is limited due to the ambiguity 
in the evolution of the various physical fields 
beyond the CH. It therefore makes sense to restrict attention to 
the object's history 
{\em up to} the CH. (Obviously, a non-catastrophic approach to the CH 
is a necessary condition for a peaceful crossing of the latter.) 
This is the approach we 
shall take here. 

There are two generic sources of electromagnetic radiation which irradiate 
the CH. First, any realistic BH  
is surrounded by the `tails' of radiation---which result from the 
backscattering off the curvature of spacetime of 
the electromagnetic waves created by the evolution of 
(electromagnetic) multipole moments in the star during the collapse. 
These `tails' decay at the event horizon according to an inverse power law 
in advanced time \cite{price,gundlach}. 
Second, any BH constantly captures photons which 
originate from the relic cosmic microwave background radiation (CBR). 
It turns out that 
both the flux and the frequency of these photons are infinitely 
blue-shifted at the CH. In this Paper we shall focus on this second 
source---the CBR photons. 

Linear analyses \cite{simpsonandpenrose} have shown that 
perturbations which outside the BH decay according to an 
inverse-power law in external time, diverge exponentially at the CH. Also,  
the projection of the energy-momentum tensor on the worldline of an 
infalling observer diverged at the CH \cite{chandrasekharandhartle}. 
Burko and Ori (BO) \cite{burkoandori} have 
recently shown that this divergence does not necessarily mean the complete 
burning up of physical objects on their attempt to traverse the CH. 
However, BO treated the electromagnetic field as a 
{\it classical} Maxwell field, 
and did not consider possible high-energy QED processes. They 
also did not discuss the other generic radiation source, namely the infalling 
photons of the CBR. Cosmological effects were 
studied by Balbinot {\it et al} \cite{balbinot1} 
who considered the traversability of a wormhole consisting of 
closed Friedmann-Robertson-Walker universes connected by a 
RN BH. They showed that the CH 
singularity was classically mild, 
but did not consider, however, the direct electrodynamic 
interaction of the CBR photons with the infalling object. 

In this Paper we analyze the (ultra-relativistic) electrodynamic 
interaction of the blue-shifted 
CBR photons with an infalling object. We consider the 
interaction up to second order in QED perturbation theory, and 
find that this interaction is dominated by the production of 
electron-positron pairs in the field of the atomic electrons and the 
nuclei of the matter comprising the object, and by the ionization 
of the matter due to the photo-electric and Compton effects.  
(We have not looked at the effects of higher-order interactions.) 
We shall show, that if the infalling object is  
small compared to the typical penetration length of the electrodynamic 
processes (i.e., if the typical length of the object in 
the radial direction is of order $1-10$ cm or smaller), only a small 
fraction of its atoms will interact with the 
infalling photons.  

As the energy of the incident photons 
diverges on the CH, the photon's energy becomes super-Planckian at some 
point. As physics at super-Planckian energies is as yet completely unknown, 
we introduce a cutoff when the photon's energy becomes Planckian. Moreover, 
physics is as yet unknown even at lower energies. We (quite artificially) 
face this difficulty by extrapolating the QED cross sections up to a cutoff 
introduced at the Planck 
energy. While this is clearly a very simplified model for the interaction 
at high energies, we hope that it may yield a reasonable order-of-magnitude 
estimate for the physical effects. This estimate may be reasonable 
if higher order QED effects and higher-energy  
interactions do not change our conclusions drastically. 
We also do not consider here semi-classical effects, like those 
studied in Refs. 
\cite{novikovandstarobinsky,poissonandmarkovic,balbinot2}.

We shall show that up to Planck energy  
the actual heating of the infalling matter will be bounded and for typical 
astrophysical parameters 
could be even quite small. Also, this heating is not 
sensitive to the cutoff energy scale. For a more complete 
understanding, one should use electroweak and GUT cross sections at 
corresponding energies, and consider QED interactions of higher order in 
perturbation theory, as well as the formation of more massive pairs and the 
back-reaction on the geometry. 
Nevertheless, we believe that our analysis can 
yield a vague order-of-magnitude estimate for the 
actual interaction strength. Finally, we note that as 
the interaction strength we calculate is 
bounded and is not drastically large, 
we do not find support to the hypothesis that there is no physical 
continuation of the geometry beyond the CH. 

The organization of this Paper is as follows: In Section II we evaluate 
the number 
of the incident photons. Then, in Section III, we assess the actual number of 
the photons interacting with the infalling object, and in Section 
IV we study the 
effect of these interaction on the object.

\section{The Number of Incident Photons}
Let us consider an isolated RN BH with mass $M$ and 
electric charge $Q_{*}$. We assume that this BH is surrounded 
by the CBR photons, which  away from the BH,    
are distributed homogeneously and isotropically, according to the standard 
Big Bang cosmological model. We also assume that the cosmology is described 
by the standard Robertson-Walker model with critical density parameter  
$(\Omega =1)$. (In fact, we find that our 
results are insensitive to the details of cosmological model.)

The flux of photons which fall into the BH is 
$\,dn(t)/\,dt=\sigma_{\rm cap}\rho (t)c,$ 
where $\sigma_{\rm cap}$ is the cross section for capture of photons 
by the BH. 
This cross section is a function of  
$Q_{*}$ and $M$, and is given explicitly in Ref. \cite{zakharov}. 
$\rho (t)$ is the 
number density of the CBR photons at the external 
time $t$  away from the BH, and $c$ is the speed of light. 
We make the simplifying assumption 
that the infalling photons move along radial trajectories described by 
geometrical optics. (This approximation seems to be justified due to  
the divergent blue-shift.) When the object gets close to the CH, the flux  
of incident photons (per unit proper time) is the (external time) 
absorption rate, $\,dn/\,dt$, multiplied by two geometric factors: First, 
by $S/(4\pi r_{-}^{2})$, where $S$ is the object's cross sectional area (in 
the $\theta - \phi$ plane), 
and $r_{-}=M-(M^{2}+Q_{*}^{2})^{1/2}$ is 
the inner horizon. (We take here $c=1$. We also take Newton's 
constant $G=1$ throughout.) 
This is the fraction of the solid angle $4\pi$ from 
which photons can hit the object. Second, the factor $\,dt/\,d\tau$, which 
relates the external time $t$ to the observer's proper time $\tau$. Here, 
the time $t$ can be taken along a line of constant $r=r_{0}\gg M$. 
[$r_{0}$ is much smaller than the typical cosmological radius of curvature. The 
relation between $t$ (at $r=r_{0}$) and $\tau$ (at the object's world line) 
is determined by ingoing null geodesics. Here, $r$ 
is the radial Schwarzschild coordinate.] A straightforward calculation 
yields $\,dt/\,d\tau\approx (\kappa_{-}\tau)^{-1}$, where $\kappa_{-}$ is 
the surface gravity of the CH defined by $\kappa_{-}\equiv \frac{1}{2}\left|
\,df/\,dr\right|_{r=r_{-}}$. Here, $f=1-2M/r+Q_{*}^{2}/
r^{2}$. (This approximation for $\,dt/\,d\tau$ is valid near the CH to leading 
order in $r-r_{-}$.)
We set $\tau$  
such that $\tau=0$ is the value of $\tau$ corresponding to the 
observer's arrival at the CH. Therefore, the flux of incident photons is 
$\dot{n}(\tau)\equiv\,dn/\,d\tau
=\rho (t)\sigma_{\rm cap}S/(4\pi r_{-}^{2})| \kappa_{-}\tau |^{-1}.$ 
Now, the variation of $\rho$ with $t$, due to the cosmological evolution, 
is unimportant to our problem: The relevant interval of $t$ (up to the 
Planckian cutoff) is negligible compared with the cosmological evolution 
time-scale. We can thus replace $\rho (t)$ by $\rho_{0}$, which is taken to be 
the density of the CBR photons when the observer jumps into the BH. In what 
follows we take $\rho_{0}$ to be the present density of the CBR photons. We 
thus obtain $\dot{n}(\tau)=\rho_{0}
\sigma_{\rm cap}S/(4\pi r_{-}^{2}) | \kappa_{-}\tau |^{-1}.$ 
The total number 
of incident photons is obtained by integrating $\dot{n}$ over $\tau$ 
from some $\tau_{\rm c}$, which is 
taken, e.g., to be the proper 
time corresponding to the threshold energy for pair production, to some final 
proper time $\tau_{\rm f}$---which will be taken later to be the proper 
time at which incident photons are blue-shifted to Planck 
energy. The total number of photons above the threshold energy hitting 
the infalling object is thus 
\begin{equation}
n(\tau_{\rm f})=\sigma_{\rm cap}\rho_{0}\kappa_{-}^{-1}
S/(4\pi r_{-}^{2})\ln \tau_{\rm c}/\tau_{\rm f}.
\label{in1}
\end{equation}

\section{The Number of Interacting Photons}
In this section we first discuss the number of the pair-production events 
and then we discuss the number of Compton scattering and photo-electric 
effect events. We shall see that the former is larger than the latter 
by a factor 
$Z$, $Z$ being the electric charge of the nucleus. (However, the contributions  
of the latter to the thermal effect we shall discuss in Section IV 
are of the same 
order of magnitude---see below.) 

\subsection{The Number of Pair-Production Events}
To obtain the actual number of photons {\em interacting} with the atomic 
nuclei or  
electrons of the infalling object, we need to multiply Eq. (\ref{in1}) by 
$1-\exp (-x\rho_{\rm ob}\sigma_{e^{+}e^{-}})$, where $\rho_{\rm ob}$ is the 
number density of atomic nuclei (electrons), $\sigma_{e^{+}e^{-}}$ is 
the corresponding 
cross section for pair production, and $x$ is the length of the object in 
the radial direction (or its thickness). 
(We use here the constancy of the cross section in 
the ultra-relativistic limit---see below.) 
Taking 
$x\rho_{\rm ob}\sigma_{e^{+}e^{-}}\ll 1$, we find that the total number of 
pair-production events per unit volume is
\begin{equation}
{\cal N}_{e^{+}e^{-}}(\tau_{\rm f})=\sigma_{\rm cap}\sigma_{e^{+}e^{-}}\rho_{0}
\rho_{\rm ob}\kappa_{-}^{-1}/(4\pi r_{-}^{2})\ln\tau_{\rm c}/
\tau_{\rm f}.
\label{in2}
\end{equation}
(In fact, our results are still valid even if we relax this assumption, as 
long as $x\rho_{\rm ob}\sigma_{e^{+}e^{-}}$ is not much 
larger than unity.)   
If the infalling object is much smaller than the typical mean free path of 
the photons, we can assess the interaction strength by calculating 
the probability that no pair-creation events will occur. 
It turns out that the majority of the interactions are those which occur in 
the Coulomb field of the atomic nuclei. 

To evaluate the probability that up to $\tau_{\rm f}$ {\em none} 
of the atomic nuclei (electrons) 
will interact with the incident photons, we note that from 
Eq. (\ref{in2}), the number of interactions per nucleus (electron) is 
\begin{equation}
{\cal P}=\sigma_{\rm cap}\sigma_{e^{+}e^{-}}\rho_{0}
\kappa_{-}^{-1}/(4\pi r_{-}^{2})\ln\tau_{\rm c}/\tau_{\rm f}.
\label{in3}
\end{equation}
(As we shall see below, ${\cal P}\ll 1$.) Therefore, the object will 
most likely arrive at the 
CH without experiencing even a single interaction if the number $\nu$ 
of its nucleons 
(electrons) is small compared to ${\cal P}^{-1}$. 

We express $\sigma_{\rm cap}$  in terms of 
the BH parameters, and find that 
$\sigma_{\rm cap}=\pi l M^{2}$, where $l$ is a known dimensionless 
function of the 
charge-to-mass ratio of the BH $q$ \cite{zakharov}.  
We also find that 
$\tau_{\rm c,f}\approx -\kappa_{-}^{-1}{\cal E}_{0}/{\cal E}_{\rm c,f}$, where 
${\cal E}_{0}$ is the photon's energy away from the BH.  

We now assume that $q=0.998$ \cite{thorne}, and obtain  
$\sigma_{\rm cap}\approx 3.5\times 10^{11}(M/M_{\odot})^{2}\;{\rm cm}^{2}$, 
$1/r_{-}^{2}\approx 5.2\times 10^{-11} (M/M_{\odot})^{-2}\;{\rm cm}^{-2}$, 
and $1/\kappa_{-}\approx 1.0\times 10^{6}(M/M_{\odot})\;{\rm cm}$. 
$\tau_{\rm c}\approx -5.0\times 10^{-15}(M/M_{\odot})\;{\rm sec}$ and 
$\tau_{\rm f}=\tau\left({\cal E}_{\rm Planck}\right)
\approx -6.4\times 10^{-37}(M/M_{\odot})\;{\rm sec}$. 
Here, $M_{\odot}$ denotes the solar mass.
 
The Bethe-Heitler formula \cite{betheandheitler} for the ultra-relativistic 
cross section for pair 
production in the Coulomb field of a light nucleus yields, in the case 
of complete screening, 
$\sigma_{e^{+}e^{-}}=\alpha r_{\rm c}^{2}Z^{2}\left[ \frac{28}{9}
\ln\left(183Z^{-\frac{1}{3}}\right)-\frac{2}{27}\right]$, where 
$\alpha$ is the fine-structure constant, and $r_{\rm c}$ is 
the classical radius of the electron. (For the corresponding expression 
for the field of the atomic electrons set $Z=1$.) 
 
We find that 
$\sigma_{e^{+}e^{-}}\approx 9.4\times 10^{-27}Z^{2}\;{\rm cm}^{2}$. 
Hence, we find that 
${\cal P}\approx 3\times 10^{-16}(M/M_{\odot})Z^{2}.$ 
Taking $M/M_{\odot}=10^{6}$, we obtain  
${\cal P}\approx 3\times 10^{-10}Z^{2}$. 

Typically, the number of interactions will be dominated by the Coulomb field of 
the nuclei and not by the field of the electrons, because 
the cross section for the former is 
proportional to $Z^{2}$, while the number of electrons per nuclei 
is $Z$. Consequently, the total number of events in the field of the nuclei 
is larger than the number of events in the field of the atomic electrons 
by a factor $Z$. 
We conclude that if the object is microscopic, with number of atoms  
$\nu <{\cal P}^{-1}\approx 3\times 10^{9}Z^{-2}$, it is most likely 
that none of the 
atomic nuclei (or electrons) will interact with the incident photons. 

A larger object will be affected by the interaction. Yet, we find 
that only a small fraction---namely, just $3\times 10^{-10}Z^{2}$---of 
its atomic nuclei will interact.  
However, such a large object might be heated by the  
energy transfer to the electrons.

\subsection{The Number of Compton and Photo-electric Events}
Ultra-relativistic photons can interact with matter also through 
ionization of the 
matter due to the photo-electric effect and the Compton effect. In the limit 
of extremely hard photons, we expect these two effects to behave 
similarly. Consequently, 
we shall discuss in detail only the latter. 

The cross section for the Compton effect in the ultra-relativistic limit 
is given by \cite{heitler}  
$\sigma_{\rm Com}({\cal E})=\pi r_{\rm c}^{2}\frac{{\cal E}_{\rm c}}{{\cal E}}
\left(\ln  \frac{2{\cal E}}{{\cal E}_{\rm c}} +\frac{1}{2}\right)$. 
The total number of events will be given by integrating the product of this  
cross section and the flux of the incident 
photons over the proper time up to the CH.  
One finds, then, that the total number of Compton scatterings per atomic 
electron is 
\begin{equation}
{\cal N}_{\rm Com}=-{\cal N}_{0}\int_{1}^{\infty}
\frac{{\cal E}_{\rm c}^{2}}{{\cal E}^{2}}\left(\ln \frac{2{\cal E}}{{\cal 
E}_{\rm c}} +\frac{1}{2}\right)\,d\left(\frac{{\cal E}}{{\cal E}_{\rm c}}
\right) ,
\end{equation}
where ${\cal N}_{0}=\rho_{0}\sigma_{\rm cap}/(4\kappa_{-})\left( 
r_{\rm c}/r_{-}\right)^{2}$. 
Here, the lower limit of the integration is taken to be the energy of 
the incident photon 
corresponding to the electron mass $m$. For the same numerical values as above, 
one finds that ${\cal N}_{\rm Com}\approx 3\times 10^{-16} (M/M_{\odot})$. 
This is of the same order as the number of pair-production events in the 
field of the 
atomic electrons. (Recall that the total number of pair-production 
events in the 
field of the nuclei is larger by a factor $Z$.)  
Note that ${\cal N}_{\rm Com}$ is 
dominated by the lower limit of the integral: 
the number of events at higher energies 
is vanishing with the increase in the energy. 
Therefore, we assume that {\em all} the 
Compton scattering events occur at energies of the same order as $m$. 

\section{The Heating Effect}

In this section we first calculate the thermal effects due to the pair-creation 
processes, which are dominated by the atomic electrons, and then we discuss 
the thermal effects due to Compton scattering. 

The energy of each interacting photon 
is extremely large, and is growing on the approach to the CH up to (and 
possibly even beyond) Planckian levels. Hence, if the size of the object 
in the radial direction is much larger than the typical mean free path 
for the relevant processes, one would expect 
that a considerable portion of the energy of the photons will be absorbed 
by the object. As this energy is very large, the inevitable result is the 
complete destruction of the object. This expectation is based on the 
scenario of  
a multiplicative shower, in which incident photons produce $e^{+}-e^{-}$  
pairs, which, in turn, produce even more photons through Bremsstrahlung 
processes. These photons are expected to create more pairs, etc. As the 
typical mean free paths for pair production and Bremsstrahlung in the 
ultra-relativistic limit are roughly equal (for water they are each of order 
$50\; {\rm cm}$), one could simplify the calculations by considering the 
effects of a single interaction for which the effective mean free path is 
$\lambda$. 
(The effects of other processes, e.g., annihilation of the created positrons 
with atomic electrons, are expected to be negligible.) For a thick object 
we find that the result we shall obtain below in Eq. (\ref{in4}) should 
be multiplied by a factor $e^{x/\lambda}$. Hence, for $x$ of the same 
order as $\lambda$, one would obtain just a numerical factor of order 
$1-10$.  

However, if the object is thin compared to $\lambda$, 
the created pair will most probably cross the object and 
leave it without any further interaction. Consequently, we shall 
discuss objects of typical thickness smaller than $\lambda$. 
(Note that we do not assume $x\ll\lambda$.) 
As the created pair does not interact with the object, 
the only source of energy which can be absorbed by the later is the recoil 
of the nuclei or of the electrons during the interaction.
Due to exchange effects the 
maximum momentum transfer in the pair production process (in the field of 
the atomic electrons or nuclei) is of 
order $m$ \cite{josephandrohrlich,landauandlifshitz}. As the momentum 
of the photon is much larger than $m$, its momentum is almost completely 
transferred to the created pair, which will consequently move 
almost in the same direction of the original photon. 

The absorbed energy is inversely proportional to the mass of the nucleus 
or the electron. Consequently, the energy absorbed by the recoiling 
electrons is much larger than the energy absorbed by the nuclei. Therefore, 
the energy absorption by the nuclei is smaller by a factor $Zm/\mu$, 
where $\mu$ is the mass of the nucleus. 
We shall 
focus, therefore, on the thermal effects due to the interaction with the 
electrons. We shall make the simplifying assumption, that the 
energy absorption is dominated by the large momentum transfer (LMT) events. 
Namely, we shall consider here only the events in which the momentum transfer 
is of order $m$. However, we then assume that in {\em all} the events the 
momentum transfer is $m$.  (This is just an order-of-magnitude estimate. 
Therefore, there is an unknown factor of order unity in our results.) 

The LMT cross-section is 
$\sigma_{e^{+}e^{-}}^{\rm lmt}=(82/27)\alpha r_{\rm c}^{2}$ 
\cite{wheelerandlamb,josephandrohrlich}.  
We then assume that all of the transferred 
energy is absorbed in the infalling body as thermal energy, which is 
manifested by the heating of the object.  
(The stopping range in biological matter for electrons with kinetic 
energy of order $0.25\; {\rm MeV}$ is less than $1\; {\rm mm}$ 
\cite{stopping}. Hence, if our object were such that $x\ge 1\; {\rm mm}$ we 
indeed find that all of the kinetic energy would be absorbed through 
all channels of energy loss by charged particles moving in matter.) 

We thus find that the thermal energy absorbed by the infalling object 
up to Planck energy is
\begin{equation} 
K\approx\left(\sqrt{2}-1\right)mc^{2}\sigma_{\rm cap}
\sigma_{e^{+}e^{-}}^{\rm lmt}\rho_{0}\kappa_{-}^{-1}/(4\pi r_{-}^{2})
\ln\tau_{\rm c}/\tau_{\rm f}.
\label{in4}
\end{equation}
Taking the numerical values for a BH with 
$M=10^{6}M_{\odot}$ and $q=0.998$ we find that the absorbed thermal 
energy per gramme is $k\approx 0.6\; {\rm J}/{\rm g}$, where we 
took water to be the matter the infalling object is made of. 
(It turns out that biological matter---as it is composed 
mainly of water---have very similar 
physical properties to water. Therefore, we simplify our calculations by 
modelling the infalling object to be made of liquid water.) This 
corresponds to an increase in the temperature of $0.1\; {\rm K}$. 

The contribution of Compton scattering to the thermal effects is expected 
to be of the same order as the contribution of the pair-production events 
calculated above: Recall that the number of Compton scatterings is of the 
same order as the number of pair productions in the field of the atomic 
electrons. As the former is dominated by incident energies of order $m$, 
we approximate the energy transferred to the ejected electron by $m$.  
Consequently, a repetition of the above analysis yields for the contribution to 
the increase in the temperature again $0.1$ K.  

Note that we did not 
consider the contribution of  interactions a {\em lower} energies. 
Indeed, the threshold 
energy for pair production sets a lower cutoff in the energy of the 
incident photon. However, there is no corresponding cutoff when Compton 
scattering or the 
photoelectric effect are concerned. Therefore, a more complete 
analysis of the effects 
experienced by the infalling object should 
include these 
effects also at energies corresponding to the non-relativistic limit 
up to $m$. In addition, it turns out that the effects caused by the 
excitation and the photo-disintegration of the nuclei are also not 
expected to be catastrophic. Consequently, we expect that the overall 
interaction 
of the infalling object with the CBR photons is bounded and small. 

We have found that the more massive the BH, the stronger the 
interaction of the blue-shifted CBR photons with the infalling object. 
Therefore, to decrease the extent of this interaction, it would be natural 
to take less massive BHs. However, 
in such BHs tidal effects could be disastrous for physical 
objects. 

In conclusion, we find that the thermal 
heating of the infalling 
object could be small enough to allow us not to exclude the 
possibility that it will arrive negligibly damaged at the CH.  However, note 
that as far as a {\em living} 
observer is concerned, this heating corresponds to 
radioactive radiation of $0.1$ K. 
{\it It should be remembered, that such a radiation would probably be 
fatal for a macroscopic living organism such as a human being}. However, 
micro-organisms may arrive at the CH without experiencing even a single 
interaction, and for them the thermal effects would consequently be 
irrelevant. A non-living object whose typical size is $1-10\; {\rm cm}$ or 
smaller might also be destroyed by this radiation unless 
it is insensitive to such radioactive radiation. 
Yet, this investigation does {\it not} provide support to the idea that the CH 
is to be regarded as a wall which cannot be traversed (or as the 
boundary of spacetime). 

\section*{Acknowledgements}
I would like to thank Aharon Bar-David, Joshua Feinberg, Yithak 
Gertner, and Norma S\'{a}nchez. 
It is especially a great pleasure for me to thank Amos Ori for many 
helpful and clarifying discussions. 


\end{document}